\documentclass[aps,prl,showpacs,eqsecnum,twocolumn,superscriptaddress]{revtex4}
\usepackage{amsmath,amssymb}
\usepackage{graphicx}

\begin{document}

\title{Gravitational waves from the Papaloizou-Pringle instability in black hole-torus systems}

\author{Kenta Kiuchi}
\affiliation{Yukawa Institute for Theoretical Physics, 
Kyoto University, Kyoto, 606-8502, Japan~} 

\author{Masaru Shibata}
\affiliation{Yukawa Institute for Theoretical Physics, 
Kyoto University, Kyoto, 606-8502, Japan~} 

\author{Pedro J. Montero}
\affiliation{Max-Planck-Institute f{\"u}r Astrophysik,
Karl-Schwarzschild-Str. 1, 81748, Garching bei M{\"u}nchen, Germany}

\author{Jos\'{e} A. Font}
\affiliation{Departamento de Astronom\'{\i}a y Astrof\'{\i}sica,
Universitat de Val\`encia, Dr. Moliner 50, 46100 Burjassot, Spain}

\date{\today}

\begin{abstract}

Black hole (BH)--torus systems are promising candidates for the
central engine of gamma-ray bursts (GRBs), and also possible outcomes of the collapse
of supermassive stars to supermassive black holes (SMBHs). By three-dimensional
general relativistic numerical simulations, we show that an $m=1$ nonaxisymmetric
instability grows for a wide range of
self-gravitating tori orbiting BHs. The resulting  nonaxisymmetric
structure persists for a timescale much longer than the dynamical one, becoming a strong emitter of large amplitude, 
quasiperiodic gravitational waves. Our results indicate that both, the central
engine of GRBs and newly formed SMBHs, can be strong
gravitational wave sources observable by forthcoming ground-based and spacecraft detectors. 

\end{abstract}


\maketitle


\emph{Introduction.}--- Black hole (BH)--torus systems are common
in the universe. The central region of active galactic nuclei is
believed to consist of a supermassive black hole (SMBH) of mass
$M_{\rm BH}\sim 10^{6}$--$10^{10}M_\odot$ surrounded by a torus. Such
systems may form through the collapse of supermassive stars
(SMS)~\cite{Rees:1984si,Shibata:2002br}. Mergers of neutron star (NS) binaries
and BH-NS binaries often result in a BH and a torus~\cite{NSNSBH}. 
Such systems can also be produced at the end of the
life of massive stars~\cite{Heger:2002by}.  
  It has been suggested that the merger and
collapsar scenarios are linked to short- and
long-duration gamma-ray bursts (GRBs), respectively 
\cite{Narayan:1992iy,Woosley:1993wj}. Thus, BH-torus systems may  become high-energy astrophysical
objects.  However, their formation and evolution have
not yet been observed as such sites are opaque to electromagnetic
waves (EWs) due to their intrinsic high density and temperature.
Gravitational waves (GWs), however, are much more transparent than EWs with
respect to absorption and scattering with matter. If BH--torus systems emitted
detectable GWs, it would be possible to 
explore their formation and evolution,  along with the prevailing hypotheses that
associate them to GRB engines.
 This may become possible in the near future thanks to the array of
ground-based (LIGO, VIRGO, LCGT) and spacecraft (LISA)
detectors of GWs~\cite{Detectors}.
 
BH-torus systems are known to be subject to the axisymmetric runaway
instability~\cite{Abramowics:1983}, and the nonaxisymmetric Papaloizou-Pringle
instability (PPI)~\cite{Papaloizou:1984}. 
Recently, general relativistic (GR) simulations have shown  
that the runaway instability does not have a significant impact on the dynamics even if
the torus self-gravity is taken into
account~\cite{Montero:2010gu}. However, nonaxisymmetric features
may still play a crucial role as shown by linear analysis~\cite{Narayan:1987zz}. Exploring the nonlinear growth and saturation of the
PPI in BH--torus systems
requires three-dimensional (3D) simulations~\cite{Blaes:1988}. Moreover, since GR plays an
important role in the case of self-gravitating
tori~\cite{Korobkin:2010qh}, 3D GR numerical simulations are mandatory. 
In particular, long-term simulations are essential to follow the BH--torus systems 
after the PPI saturation. By contrast, the recent 3D GR simulations 
of BH--torus by Korobkin et al. 
only followed the linear growth phase of the PPI~\cite{Korobkin:2010qh}.

In this {\it Letter} we report results of a sample of such simulations
which show that the PPI sets in for a wide range of BH--torus systems, and  that the resulting
nonaxisymmetric structure is maintained after the saturation of the instability.
As a result, such systems can be strong sources of detectable GWs
for the detectors that will be in
operation in the near future.  We use units
of $c=G=1$, and the perturbation is expressed by the expansion of ${\rm
exp}\{-i(\omega t - m \varphi)\}$.

\emph{Method and initial models.}--- The simulations are
performed using a fixed mesh refinement (FMR) and 
Message Passing Interface version of the SACRA code (see~\cite{Yamamoto:2008js} for details regarding  the implementation of the FMR algorithm and the
formulation and numerical schemes for solving Einstein's and
hydrodynamics equations). The FMR structure consists of 6 refinement
levels for which  the grid resolution changes as $\Delta x_{l-1}=2\Delta
x_{l}$ ($l=2$--6), with a finest level of $\Delta x_{l=6}=0.05
M_{\rm BH}$.  Each domain is composed of $(2N+1, 2N+1, N+1)$ Cartesian
grid zones $(x, y, z)$ which cover the interval $[-N\Delta x_l:N \Delta x_l]$ for the
$x-$ and $y-$directions and $[0:N \Delta x_l]$ along $z$, with $N=98$.  The outer
boundary is located at $156.8M_{\rm BH}$ which is in the local wave zone
(see Table~\ref{tab:model}). The BH horizon is covered by $\approx 10$
grid points in the finest refinement level. For assessing the validity
of the numerical results, simulations with different values of $\Delta
x_{l=6}$, $N$, and size of each domain were performed, which
confirmed that convergence is achieved with our chosen values.  We use a standard
$\Gamma$-law equation of state (EOS) $P=(\Gamma-1)\rho\varepsilon$,
where $P$ is the pressure, $\rho$ the rest-mass density, $\varepsilon$
the specific internal energy, and $\Gamma$ the adiabatic index.

\begin{table*}
\centering
\begin{minipage}{140mm}
\caption{\label{tab:model} List of key quantities for the initial
  data: specific angular momentum profile (value of $j/M_{\rm BH}$
  at the inner edge), torus-to-BH mass ratio, ${\cal R} \equiv M_{\rm  tor}/M_{\rm BH}$, 
position of the inner edge, center (location of
  the maximum density), and outer edge on the equatorial plane,
  $r_{\rm in}$, $r_{\rm c}$, and $r_{\rm out}$, orbital period at
  $r=r_{\rm c}$, $t_{\rm orb}$, and simulation time in units of
  $t_{\rm orb}$. The last two columns show the growth rate
  Im($\omega_m$) of the PPI and the Fourier amplitude of the density
  perturbation for the $m=1$ mode at the saturation, $\delta_1 \equiv
  |D_1|/D_0$. Here, $\Omega_{\rm c}$ is the angular velocity at
  $r=r_{\rm c}$.  }
\begin{tabular}{llcccccccc}
\hline\hline
Model~~~~                     &
$j$ profile                   &
$M_{\rm tor}/ M_{\rm BH}$  &
$r_{\rm in}/ M_{\rm BH}$    &
$r_{\rm c} / M_{\rm BH}$    &
$r_{\rm out}/ M_{\rm BH}$   &
$t_{\rm orb}/ M_{\rm BH}$   &
$t_{\rm sim} / t_{\rm orb}$ &
${\rm Im}(\omega_1)/\Omega_{\rm c}$   &
$\delta_1$                 \\
\hline
C1  & constant $(3.83)$        & 0.10& 3.3&8.02&19.8& 172.6 &
$\approx$ 34       & 0.142 
& $\approx$ 0.1 \\
C06 & constant $(3.80)$        & 0.06& 3.6&8.01&19.8& 171.6 &
$\approx$ 40       & 0.089 
& $\approx$ 0.1 \\
NC1 & $j\propto r^{0.26}(3.53)$& 0.10& 4.8&10.9&19.8& 257.9 &
$\approx$ 36       & 0.056 
& $\approx$ 0.25\\
NC06& $j\propto r^{0.29}(3.62)$& 0.06& 5.0&11.7&20.0& 284.0 &
$\approx$ 34       & ----- 
& ----          \\
\hline\hline
\end{tabular}
\end{minipage}
\end{table*}

Compact equilibria for a BH--torus system are prepared in the puncture
framework and used as initial conditions~\cite{Shibata:2007zzb}. We adopt the
$\Gamma=4/3$ polytropic EOS, $P=\kappa\rho^{\Gamma}$, to mimic the EOS
of a degenerate relativistic electron gas or radiation dominated gas.
The polytropic constant $\kappa$ is chosen such that the
torus-to-BH mass ratios are in the range 
${\cal R}=M_{\rm tor}/M_{\rm BH}=0.06$--0.1. 
We specify the angular momentum profiles,
$j=j(\Omega)$, where $j\equiv (1+\varepsilon+P/\rho) u_{\varphi}$ 
and $\Omega=u^{\varphi}/u^t$, $u^{\mu}$
denoting the fluid four velocity. Table~\ref{tab:model} lists key
quantities of the four different BH-torus systems considered. The BH is assumed to
be non-rotating. We investigate both $j=$constant (C
models), and non-constant profiles (NC models) with $j \propto \Omega^{-1/6}$, to assess their influence. 
The selected mass ratios are partially motivated by the 
result of simulations of gravitational collapse to a BH and
torus~\cite{Sekiguchi:2010ja,Shibata:2002br}. The numerical model
of~\cite{Sekiguchi:2010ja} may be regarded as a candidate of the
central engine of GRBs, and that of~\cite{Shibata:2002br} as a model
for the formation of a SMBH through the collapse of a rotating SMS.

Our numerical simulations are performed for a time long enough, i.e.~$t \sim
20$--$40~t_{\rm orb}$, to include the growth, saturation, and
post-saturation phases of the PPI, for which the growth timescale
is of order $t_{\rm orb}$ (see Table I). 

\begin{figure}[t]
\centerline{
\includegraphics[width=8.0cm,angle=0]{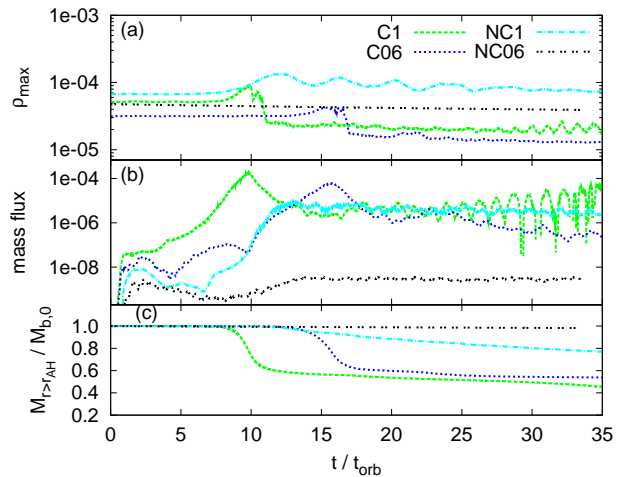}
}
\vspace{-3mm}
\caption{\label{fig1} Evolution of (a) the maximum rest-mass density, (b) the
  mass flux on the BH horizon, and (c) the rest-mass located
  outside the apparent horizon (AH) for each model. The maximum
  density is evaluated outside the AH and rest-mass is normalized by
  its initial value.  }
\end{figure}

\begin{figure}[t]
\centerline{
\includegraphics[width=8.0cm]{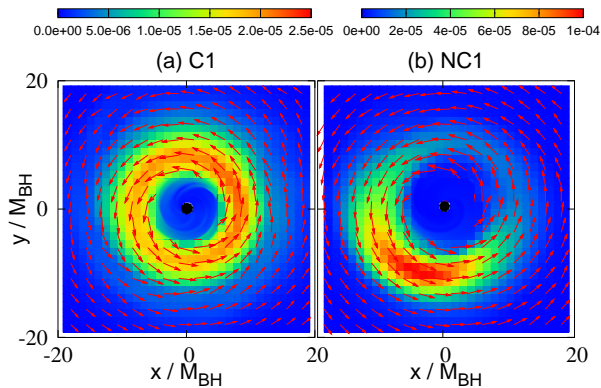}
}
\vspace{-3mm}
\caption{\label{fig2}
Snapshots of the rest-mass density distribution 
on the equatorial plane for models C1 at $t=15.11$ $t_{\rm orb}$ (left)
and NC1 at $t=20.02$ $t_{\rm orb}$ (right). 
Vectors show the velocity fields, and black filled circles around 
the centers show the region inside the AHs. 
}
\end{figure}

\begin{figure}[t]
\centerline{
\includegraphics[width=8.0cm]{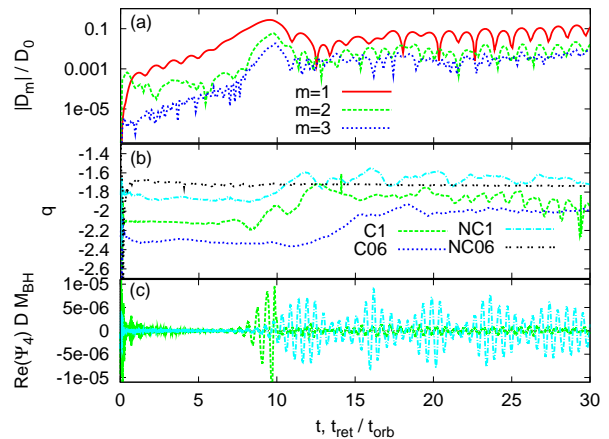}
}
\vspace{-3mm}
\caption{\label{fig3} Evolution of (a) the $m=1$--$3$ mode amplitudes for
  model C1,  (b) the power--law index of the angular velocity profile
  for all models, and (c) the outgoing part of the complex Weyl scalar
  Re($\Psi^{(2,2)}_4$) for models C1 (green) and NC1 (blue).  }
\end{figure}

\emph{Results.}--- Figure~\ref{fig1} shows the evolution of the maximum
baryon rest-mass density of the torus, the baryon rest-mass flux
measured on the apparent horizon (AH), $\dot{M}_{\rm r=r_{AH}}$, 
and the total baryon rest-mass 
located outside the AH for each model. Figure~\ref{fig2}
shows snapshots of the rest-mass density distribution on the equatorial plane after the PPI saturation for
models C1 and NC1. Figure~\ref{fig3} shows the time evolution of the density
perturbation amplitudes
for the $m=1$--3 modes of model C1, a typical power-law index of the
angular velocity, and the outgoing component of the complex Weyl
scalar $\Psi^{(2,2)}_4$ with $l=m=2$ mode for models C1 and NC1, 
where $t_{\rm ret}$ is the retarded time.

These figures reveal that the PPI is present in all models except NC06. The instability
grows exponentially with time, with the
$m=1$ mode being the fastest growing mode (cf.~Fig.~\ref{fig3}(a)).
Figure~\ref{fig1}(a) indicates that during the linear growth phase, the
maximum density is approximately constant, but after several orbits,
nonlinear growth sets in and $\rho_{\mathrm{max}}$
increases. After saturation, $\rho_{\mathrm{max}}$ decreases and eventually approaches a new
quasiequilibrium value. For a given angular momentum profile, the
PPI growth timescale is shorter  the larger the mass ratio
${\cal R}$.  For a given mass ratio, the growth timescale for NC
models is longer than for C models. 

As shown in Fig.~\ref{fig1}(b), the mass flux increases exponentially with the 
PPI growth.  It reaches its peak value at mode growth saturation and then settles 
to an approximately non-zero constant value. These features are
qualitatively identical for the PPI unstable models.  The mass
flux for NC06 is negligible throughout the evolution. These features
are reflected in Fig.~\ref{fig1}(c): Around the saturation time of the
PPI nonlinear growth, the torus mass steeply decreases, and then
the torus relaxes to a stationary accretion state. For NC models, the
mass flux at the saturation time is always lower than for C models, which indicates 
the dependence  of the accretion history on the profile of $j$. 
In Fig.~\ref{fig2} we illustrate the nonaxisymmetric morphological features present
 in the tori of models C1 and NC1 once the stationary accretion phase is 
 reached. The presence of the  $m=1$ mode is apparent. These results imply that 
 more massive tori and/or a $j=$ const profile favor the appearance of the PPI with
 respect  to less massive tori and/or a $j\not=$ const
profile, for which the torus may be PP-stable  (e.g.~model NC06). In addition, no sign of
the runaway instability is found for any models.

Figure~\ref{fig3}(a) plots the evolution of the Fourier modes for the density perturbation of
model C1, for which we define $D_m=\int \rho \, {\rm e}^{-im\varphi}d^3x$ and
$\delta_m \equiv |D_m|/D_0$.  This shows that the $m=1$ mode is the
fastest growing mode and that saturation occurs at $t \sim 10$ $t_{\rm
orb}$.  We determine the PPI growth rate through a fit of the  
numerical data, and find it in a wide range $\sim 5$--25\% of the
angular velocity at the saturation time (see Table~\ref{tab:model}). As expected from 
Fig.~\ref{fig1}, massive and/or $j=$ const models exhibit larger growth rates (in agreement 
with~\cite{Korobkin:2010qh}) and no evidence of the instability is found for model NC06.

The PPI activates the transport of angular momentum outwards through the
corotation point resonance~\cite{Narayan:1987zz}, and therefore, redistributes
angular momentum. Figure~\ref{fig3}(b) shows the evolution of a typical power-law index of
the angular velocity profile, where we fit $\bar{\Omega}(r) \equiv 1/2\pi
\int  \Omega(r,\pi/2,\varphi) \,d\varphi$ by $\propto r^q$ for $r \in
[r_{\rm in},r_{\rm out}]$ on the equatorial plane.  In Newtonian
gravity $q=-2$ for $j=$ const and $q=-1.5$ for Keplerian motion. It is
expected that $q$ would approach $-1.5$ (Keplerian limit) from a lower
initial value $< -1.5$. Note that in GR $j=$ const law
does not correspond to $q=-2$ and also $q \not=-1.5$ for the Keplerian limit.
Figure~\ref{fig3}(b) shows that after the onset of the PPI (for all models except
for NC06), $q$ increases and then saturates, reflecting the outward
transport of angular momentum.  The increase of $q$
indicates that tori with a positive value of $dj/dr$ are stable against
the PPI, as expected from linear analysis~\cite{Narayan:1987zz} and
from previous nonlinear simulations~\cite{Blaes:1988}.  All the
features found here suggest that the PPI plays an important role for
the BH-torus system dynamics unless the torus mass is small and the angular momentum 
profile is close to the Keplerian profile.

A point worth stressing is that the $m=1$ nonaxisymmetric structure
survives with an appreciable amplitude after the saturation of the
PPI nonlinear growth, as shown in Fig.~\ref{fig2}.  The likely reason 
is that as the common center of mass for the BH--torus system should remain
at the origin and the BH center is not located at the center of
mass, then the torus should be deformed. The persistent $m=1$
structure leads to the emission of  quasiperiodic GWs with a large
amplitude. 
Note that the dominant mode of GWs is $l=m=2$ because the situation is similar to a 
test particle orbiting around a BH. 
An outgoing component of the Weyl scalar
extracted in the local wave zone, plotted in Fig.~\ref{fig3}(c) for
model C1 (green curve), shows that the waveforms reflect the evolution of the
system: A burst at $t_{\rm ret}/t_{\rm orb}\approx 10$ is emitted by
the PPI nonlinear growth and saturation, and subsequent
quasiperiodic waves for $t_{\rm ret}/t_{\rm orb}\gtrsim 10$ are
emitted by the long-lived $m=1$ structure.  This feature is universal
for all the C models. While the GW for the NC1 model (Fig.~\ref{fig3}(c), blue curve) 
does not show a prominent burst, they do show the quasiperiodic emission
induced by the long-lived $m=1$ structure. The striking modulation in the GW signal found
only in $j\not=$ const models may be
due to the variability in the maximum rest-mass density and the $q$-index for the angular velocity after mode growth saturation (between $10 t_{\rm orb} < t < 30 t_{\rm orb}$ for model NC1). Therefore, it may be associated with the excitation of $p$-modes (pressure-inertial modes) in the torus. These  $p$-mode oscillations are gradually damped, as well as the modulation, for $t>30t_{\rm orb}$. The emission of quasiperiodic GWs is expected to continue as far as
the disk has a significant amount of mass. Such time duration can be estimated from
the mass accretion rate, and the residual disk mass
(Fig.~\ref{fig1}(b) and (c)). We find the accretion timescales are $t_{\rm acc} \sim M_{\rm r>r_{AH}} /
\dot{M}_{\rm r=r_{AH}} \approx 1$--$8\times 10^4$ $M_{\rm BH}$.

\begin{figure}[t]
      \includegraphics[width=8.cm]{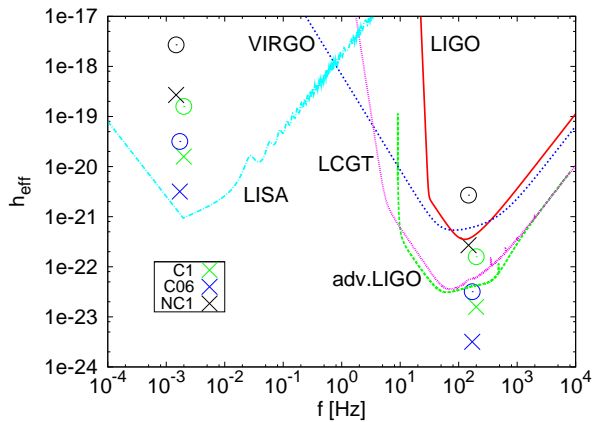}
\caption{\label{fig4}
GW spectra for all the models for
$M_{\rm BH}=10M_\odot$ and $D=100$~Mpc, and
$M_{\rm BH}=10^6M_\odot$ and $D=10$~Gpc. 
The cross symbols show the peak amplitudes found in the 
numerical simulations and the circles show hypothetical 
amplitudes inferred from accretion timescales (see main text for details).
}
\end{figure}

In Fig.~\ref{fig4} we show with crosses the peak effective amplitude $h_{\rm eff}$ of
GWs together with the design sensitivities of ground-based and spacecraft GW detectors~\cite{Detectors}. Here
$h_{\rm eff}=(\{|\tilde{h}_+|^2+|\tilde{h}_\times|^2\}/2)^{1/2}D/M_{\rm BH}$
with $\tilde{h}_{+,\times}$ and $D$ being the Fourier component of 
two polarization modes and the distance from the source.  Since our
focus is on the GW emission after the PPI growth saturation, we
discard the waveform before the saturation for the C models, 
e.g. for $t_{\rm ret}/t_{\rm orb}\lesssim 10$ for C1, when calculating $h_{\rm eff}$.  
From the accretion timescale $t_{\rm acc}$, we may expect
that the GW emission will continue for $N_{\rm cycle} \sim t_{\rm
acc}/t_{\rm GW} \approx 100$ where $t_{\rm GW}$ is the typical
oscillation period of GWs. Hence, the peak amplitudes may be enhanced
by $\sqrt{N_{\rm cycle}} \sim 10$, values plotted with circles in
Fig.~\ref{fig4}. Because we use units of $M_{\rm BH}=1$
in our simulations, the numerical results can be rescaled for
arbitrary BH mass. We choose the hypothetical values for the collapsar
model of GRB central engine as $(M_{\rm BH},D)=(10M_\odot,100~{\rm
Mpc})$ and for the hypothetical formation scenario of a SMBH through
SMS collapse as $(M_{\rm BH},D)=(10^6M_\odot,10~{\rm Gpc})$ in
Fig.~\ref{fig4}.

For the stellar-mass BH, $h_{\rm eff}$ could be $\sim 10^{-21}$ at the
peak with the frequency 100--200~Hz, which corresponds to the Kepler
frequency of $r_{\rm orb} \approx 10 M_{\rm BH}$. The peak amplitudes
agree with the order estimation, 
\begin{align*}
&h_{\rm peak}
\sim \frac{2 M_{m=1} v^2}{D}\sqrt{N_{\rm cycle}} \nonumber\\
&\approx 3 \times 10^{-21} \left(\frac{\delta_1}{0.1}\right)
\left(\frac{{\cal R}}{0.1}\right)
\left(\frac{M_{\rm BH}}{10 M_\odot}\right)
\left(\frac{100{\rm Mpc}}{D}\right)
\left(\frac{N_{\rm cycle}}{100}\right)^{1/2}
\end{align*}
where $M_{m=1}$ is the mass of matter which contributes to the $m=1$
structure (see also Table~\ref{tab:model}) and we assume the Kepler
motion with radius $\approx 10 M_{\rm BH}$.  The amplitude of the
enhanced peaks could be larger than the noise level of the advanced
ground-based detectors.  For the SMBH formation scenario with $(M_{\rm
BH},D)=(10^6M_\odot,10~{\rm Gpc})$, the predicted peak amplitude is
$h_{\rm eff} \sim 10^{-18}$--$10^{-19}$ with the frequency $\sim
10^{-3}$~Hz. Such GWs would be detected with a high
signal-to-noise ratio $\agt 10$ by LISA.

\emph{Summary and discussion.}--- We have explored the PPI in BH--torus
systems in the framework of numerical relativity. We have found that the
$m=1$ PPI occurs for a wide range of self-gravitating tori
orbiting BHs. The resulting $m=1$ structure in the torus survives
for a long time after the PPI growth saturation, and large amplitude,
quasiperiodic GWs are emitted.  For reasonable
hypothetical masses and distances, we have found that the
quasiperiodic GWs emitted by this mechanism could be a promising source 
for ground-based  and spacecraft detectors (for stellar-mass BHs and SMBHs, 
respectively). 

The BH--torus system composed of a stellar-mass BH is a
promising candidate for the central engine of GRBs, and  our results
suggest that the so-called collapsar hypothesis~\cite{Woosley:1993wj}
may be verified via observation of GWs. In GRBs, jets are likely to
be ejected along the rotation axis of a spinning BH and torus.  If the
BH--torus system formed in the collapsar is axially symmetric, GWs of
$l=2$ and $m=0$ mode are primarily emitted in the direction
perpendicular to the rotation axis (e.g.~\cite{Sekiguchi:2010ja}). By
contrast, the quasiperiodic GWs studied in this letter are emitted along
the rotation axis because the $l=m=2$ mode is dominant.  Therefore, it
could be possible to observe GRBs and GWs simultaneously, and to explore
the collapsar scenario via GW observation. On the other hand,
in the formation of a SMBH following SMS collapse, the torus-to-BH mass
ratio is predicted to be 0.05--0.1~\cite{Shibata:2002br}.  Our results
suggest that the PPI could grow in such a system.  This implies that
this scenario may be confirmed by the GW detector LISA.


\emph{Acknowledgments.}--- KK thanks to Y. Suwa for fruitful
comments.  Numerical computations were performed on XT4 at CfCA of NAOJ
and on NEC-SX8 at YITP of Kyoto University.  This work was supported
by Grant-in-Aid for Young Scientists (B) 22740178, for Scientific
Research (21340051), for Scientific Research on Innovative Area
(20105004), by HPCI Strategic Program of Japanese MEXT, by
the Deutsche Forschungsgesellschaft (DFG SFB/Transregio 7) and by the Spanish
MICINN (AYA 2010-21097-C03-01).



\end{document}